\documentclass[usenatbib]{mn2e}
\usepackage{graphicx}
\begin{document}

\title[Fundamental parameters of RR Lyrae stars. II]
{Fundamental parameters of RR Lyrae stars from multicolour photometry
and Kurucz atmospheric models -- II. Adaptation to double-mode stars}
\author[S. Barcza and J. M. Benk\H o]
{S. Barcza
and J. M. Benk\H o\thanks{E-mail: barcza@konkoly.hu, benko@konkoly.hu}
\thanks{Guest observers at Teide Observatory, Instituto de Astrofisica
de Canarias}\\
Konkoly Observatory, PO Box 67,
                   1525 Budapest, XII,
                   Hungary
}
\date{}

\maketitle

\begin{abstract} 
Our photometric-hydrodynamic method is generalized to determine 
fundamental parameters of multiperiodic radially pulsating stars.
We report 302 $UBV(RI)_C$ Johnson-Kron-Cousins observations 
of GSC 4868-0831. Using these and published
photometric data of V372 Ser, we determine the
metallicity, reddening, distance, mass, radius, equilibrium luminosity
and effective temperature.
The results underline the necessity of using multicolour photometry,
including an ultraviolet band, to classify the subgroups of RR Lyrae 
stars properly. Our $U$ observations might reveal that GSC 4868-0831
is a subgiant star pulsating in two radial modes and that
V372 Ser is a giant star with size and mass of
an RRd star.
\end{abstract}
\begin{keywords} hydrodynamics 
-- stars: atmospheres 
-- stars: fundamental parameters  
-- stars: individual: GSC 4868-0831
-- stars: individual: V372 Ser 
-- stars: variables: RR Lyrae.   
\end{keywords}

\section{Introduction}

As described in the first paper of this series (\citealt{barc4},
hereafter Paper I)
a new method can be used to determine fundamental
parameters of RR Lyrae (RR) stars 
using broad-band optical
photometry and the conservation laws of mass and momentum in the
pulsating atmosphere. 

The first version of the method 
\citep{barc2,barc5} used the law of momentum conservation in the
frame of a uniform atmosphere approximation (UAA), that is, the
pulsation of the atmosphere is taken into account as if the
atmosphere were a rigid shell. The available Johnson-Cousins
$UBV(RI)_C$
photometries of SU Dra and T Sex \citep{barc0,barc5} were 
processed as examples because these uniformly cover 
the whole cycle of pulsation and allow a solution of the
Euler equation of hydrodynamics for the mass
${\cal M}_{\rm a}$ 
of the star and distance
$d$
to it. (The subscript 'a' indicates that this mass is a  
dynamical mass derived from an analysis of the motion
of the atmosphere.)

In Paper I, an extended hydrodynamic treatment, in which the
UAA is dropped, was reported.  
The following two main steps were involved.
\begin{itemize}
\item[(i)]
Assuming a perfect spherical symmetry of the pulsation,
the photometric quantities (colour indices and brightness)
were converted to time-dependent physical quantities (effective
temperature
$T_{\rm e}$,
effective gravity
$g_{\rm e}$
and angular radius
$\vartheta$)
using the computed colours and fluxes of the ATLAS models
of \citet{kuru1}. 
\item[(ii)]
The physical quantities were introduced in the differential 
equations expressing the laws of mass and momentum conservation
during the pulsation. Particular 
solutions were given to describe the motion of the pulsating  
atmosphere in the gravity field of the star. The two time-independent 
parameters of the solutions -- the mass
${\cal M}_{\rm a}$
of the star and the distance
$d$
to it -- were determined. 
\end{itemize}

Because the ATLAS models apply to the atmosphere of non-variable stars,
quantitative photometric and hydrodynamic conditions (Conditions I
and II in Paper I, hereafter 
${\rm C}^{\rm (I)}$ and 
${\rm C}^{\rm (II)}$,
respectively) were formulated for the applicability of the quasi-static
atmosphere approximation (QSAA) in order to find the time intervals 
of the pulsation when dynamical
phenomena have a negligible effect on the colours and brightness,
(i.e. the structure and colours of the atmosphere are identical 
to those of a selected ATLAS model). 

A summary of the conditions is as follows. 
${\rm C}^{\rm (I)}$
is satisfied if the
difference of the continuum fluxes of the observed and selected ATLAS
model does not exceed the error of the observation in the optical 
spectrum covered by the colours
$U$,...,$I_C$.
${\rm C}^{\rm (II)}$ 
is satisfied if the acceleration in the atmosphere is
equal to the instantaneous 'effective gravity' 
$g_{\rm e}(t)$ 
\citep{ledo1} of the selected ATLAS model.

The 
$UBV(RI)_C$
photometry of the RRab star SU Dra was used to demonstrate that
the 
extended
method is a viable alternative to determine the 
fundamental parameters of RR stars. The atmospheric metallicity [M],
the reddening
$E(B-V)$
towards the star, 
$d$
and 
${\cal M}_{\rm a}$
were determined from phases when the conditions of
the QSAA were satisfied. 

Double-mode (DM) RR (RRd) stars pulsate in two radial modes
simultaneously. Their importance for stellar pulsation theory is
obvious because DM pulsation offers a unique possibility to
determine fundamental parameters such as mass
${\cal M}_{\rm p}$
and luminosity
$L_{\rm p}$ 
from the Petersen diagram, 
(i.e. from frequencies that are
accessible by observing the brightness variation over a sufficiently
long time scale).
${\cal M}_{\rm p}$
and 
$L_{\rm p}$ 
can be compared with the mass
${\cal M}_{\rm ev}$
and luminosity
$L_{\rm ev}$
derived from stellar evolution theory. The assumptions
${\cal M}_{\rm ev}={\cal M}_{\rm p}$,
$L_{\rm ev}=L_{\rm p}$
plus some colour information have given, 
for example, the fundamental parameters
of BS Com \citep{deka1}.

In this paper, we use our combined
photometric-hydrodynamic method to determine the fundamental
parameters 
$d$
and
${\cal M}_{\rm a}$,
the approximate position in a theoretical Hertzsprung-Russell diagram (HRD),
the radius variation, the reddening and the metallicity of the DM 
pulsators, GSC 4868-0831 and V372 Ser. We also describe the 
kinematic behaviour of the pulsating atmosphere 
in our limited hydrodynamic treatment. The accuracy of 
the light and colour curves cannot be enhanced by folding, and therefore
we give a refinement of the technique
described in Paper I. The method can be applied to RR stars with any 
number of periods
(one or $\ge 2$)
if there are photometric observations available in 
sufficient numbers. However, an application to, for example,
multiperiodic $\delta$~Sct stars with small amplitudes would allow
the determination of the 
${\cal M}_{\rm a}d^{-2}$
only because the hydrodynamic
status of the atmosphere has only a small, non-radial variation.
 
${\cal M}_{\rm a}$ 
derived here provides a 
mass value from a completely different astrophysical input
in comparison with 
${\cal M}_{\rm ev}$ 
or
${\cal M}_{\rm p}$.
Consequently, 
${\cal M}_{\rm a}$
can be an
independent check for evolution and pulsation theory of RR stars.
To the best of our knowledge, we are the first to attempt to 
determine the distance and mass, etc. of DM pulsators with an 
astrophysical method using only the motion of the atmosphere.

In comparison with the Baade-Wesselink (BW) method,
the main advantage of our method is 
that more output is obtained for less observational input.
Spectroscopic
observations are not necessary at all, and consequently our method
can easily be applied to faint stars. A BW solution has not been found 
for RRd stars in the literature. Perhaps this
can be explained by the problems arising from the faintness 
($V > 10.5$~mag for the known DM stars, \citealt{wils1}, 
\citealt{szcz1}) and multiperiodic character of DM pulsation:
simultaneous observation of light, colour curves and
spectroscopy of faint stars would be necessary over days.
Furthermore, problems are encountered with the
precise determination of the centre-of-mass velocity,
a substantial point of the BW analysis (Paper I). 
Our extended
photometric-hydrodynamic method promises to deliver parameters 
in addition to distance, the main fundamental parameter that
can be acquired by the BW analysis.

A recent challenge to pulsation theory originates from the
MOST satellite, which found frequencies of the RRd star, AQ Leo, 
with amplitudes down to the mmag level \citep{grub1}. Our method 
is an extension of the research methodology of DM stars
beyond theoretical and empirical methods using only frequencies
and amplitudes, for which data can be obtained from a single-band
time series.

The observations, standard
$UBV(RI)_C$
magnitudes of the variables and some field stars
are reported in Section 2. The metallicity and reddening of
the variables and comparison stars
are given in Section 3. As a by-product,
$T_{\rm e},\log g$,
angular radius
$\vartheta$
are also determined for the comparison stars.
In Section 4, we describe some technical details beyond those 
reported in Paper I. The results are presented here for the brightest DM
pulsators GSC 4868-0831 and V372 Ser, and an insight is given into
the kinematics of their atmospheres. We give a
discussion and our conclusions
in Sections 5 and 6, respectively.
In Appendix we describe the publicly available program package
BBK\footnote{The program package is
available from {\sl http://www.konkoly.hu/staff/barcza/pub.html}.
It is composed of tables extracted from 
{\sl http://kurucz.harvard.edu/grids.html}, 
FORTRAN source codes and a manual.} 
which can be used to determine the fundamental parameters 
from the photometric input.

\begin{table}
  \caption{Log of the observations of GSC 4868-0831.}
\label{tab1}
\begin{tabular}{lrl}
\hline
HJD$-$2\,400\,000 &  No. of frames  & Telescope\\

\hline
54822.5555-.5868 &  35  & RCC  \\
54829.4534-.6300 & 225  & RCC  \\
54830.4442-.4950 &  75  & RCC  \\
54831.4606-.6221 & 215  & RCC  \\
54832.4418-.6122 & 205  & RCC  \\
54863.4483-.7015 & 265 &   IAC80\\
54871.4757-.6242$^\ast$ &  70 & IAC80\\
54873.3661-.5941 & 235 &   IAC80\\
54874.4437-.4620$^\ast$ & 20 &  IAC80\\
54876.4205-.6108$^\ast$ &  165   & IAC80\\
\hline
\end{tabular}
\begin{list}{}{}
\item[$^\ast$] Epoch of the tie-in observations
\end{list}

\smallskip

\end{table}

\section{The observations and reduction}

\begin{figure}
  \includegraphics[width=84mm]{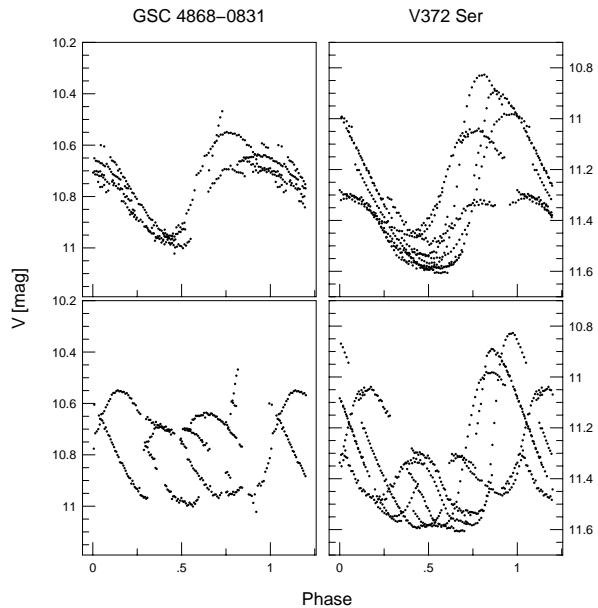}
\vspace{-0.8cm}
\caption{$V$ light curves from our observations.
The upper and lower panels are folded with
$P_1$
and
$P_0$,
respectively. 
  }
\label{fig0}
\end{figure}

The two brightest DM pulsators are the subject of the this
paper. The observational data of GSC 4868-0831 were collected with the 
IAC80{\footnote{The 0.82m IAC80 Telescope is operated on the island 
Tenerife by the Instituto de Astrofisica de Canarias in the Spanish
Observatorio del Teide.}} telescope of the Teide Observatory and 
the 1-m RCC telescope mounted at Piszk\'estet\H{o} Mountain Station of 
the Konkoly Observatory. The technical description of the CCD cameras
and the details of the observations and reductions are
identical to those of \citet{barc1}. We used the standard 
{\sc iraf}{\footnote{{\sc iraf} is distributed by the 
National Optical Astronomical Observatory,
operated by the Association of Universities for Research in Astronomy
Inc., under contract with the National Science Foundation.}} 
tasks for the reductions. 
Transformation into the standard
$UBV(RI)_C$
system was done by using the equatorial stars of \citet{land1}.
The observations of V372 Ser have been published in \citet{barc1}.
A log of the observations of GSC 4868-0831 is given in Table~\ref{tab1}, 
typical exposure times were 
180, 40, 20, 6, 10~s for
$U, B, V, (R, I)_C$,
respectively. The folded 
$V$
light curves are plotted in Fig.~\ref{fig0}.

\begin{table}
  \caption{Result of the photometry giving the average
	   values of the two variables and some comparison stars in
           the field of GSC 4868-0831. The magnitude errors of the
	   comparison stars are 
           0.008, 0.011, 0.007, 0.009, 0.009~mag in 
$V$,$U-B$,$B-V$,$V-R_C$,$V-I_C$,
respectively.}
\label{tab2}
\begin{tabular}{lrrrrr}
\hline
ID & $V$  & $U-B$ & $B-V$ & $V-R_C$ & $V-I_C$  \\
\hline
V372 Ser$^\dag$ & $11.350$ & $0.000$ & $0.380$ & $0.256$ & $0.524$  \\
GSC 4868 &&&&& \\
$-0831^{\ddag}$ & $10.769$ & $-0.087$ & $0.357$ & $0.215$ & $0.478$ \\
 &&&&& \\
$-0063$   & $12.504$ & $-0.12$ & $0.468$ & $0.287$ & $0.583$ \\	
$-0436$   & $12.637$ & $0.220$ & $0.648$ & $0.365$ & $0.705$ \\
$-0779^\ast$   & $12.534$ & $0.092$ & $0.554$ & $0.331$ & $0.641$ \\
$-0860^\ast$   & $12.579$ & $0.045$ & $0.496$ & $0.278$ & $0.519$ \\	
$-1089^\ast$   & $12.862$ & $0.073$ & $0.594$ & $0.353$ & $0.673$ \\
\hline
\end{tabular}
{\it Notes.} 
\begin{list}{}{} 
\item[$^{\dag}$]: The magnitude averaged value 
from $N=529$ observations.
\item[$^{\ddag}$]: The magnitude averaged value 
from $280$ observations.
\item[$^\ast$]: The stars are visible only in the
frames taken with IAC80.
\end{list}

\smallskip

\end{table}

The result of the tie-in observations for the comparison and check stars 
of V372 Ser is given in table~2 of \citet{barc1} and those
for GSC 4868-0831 are given in Table~\ref{tab2}. 
Additionally, we give the magnitude averaged
$V$
and colour indices of the variables obtained from our observations.
They characterize the variables because the distribution of the observations 
is quasi-random over a long enough time. 

\subsection{GSC 4868-0831} 

The 
$X=U,B,V,R_C$ and $I_C$ 
magnitudes were obtained by differential photometry
from the instrumental 
$\Delta x=\Delta u, \Delta b,...$
magnitude differences of GSC 4868-0831, GSC 4868-0063 and
GSC 4868-0436 
as follows. The coefficients 
$c_X\approx 1$
and zero points 
$c_0^X\approx 0$
of the transformation equations 
\begin{equation}\label{2.10}
X_{\rm GSC 4868-0831}=X_{\rm comp}+c_X\Delta x+c_0^X
\end{equation}
were determined from all frames separately for the telescopes 
IAC80 and RCC, respectively.
An average was computed from the comparison stars
GSC 4868-0063, GSC 4868-0436. 
Next, the magnitudes 
$U,B,R_C$
and
$I_C$ 
were interpolated to the epoch of
$V$
and the colour indices were computed.
These are the data that form the basis for our analysis.\footnote{The 
table containing 
$V$, $B-V$, $U-B$, $V-R_C$, $V-I_C$ 
observations is available electronically:
http://www.konkoly.hu/staff/benko/pub.html. 
Flag d indicates the observations
omitted from the analysis because of poor sky conditions.}

\subsection{V372 Ser}

We carefully revised the zero points 
$A_{00}^{(X)}$ 
of the light curves of V372 Ser given in table 4 of \citet{barc1} 
because it is particularly important to have 
the colour indices in the standard system. An error in 
the zero point of the magnitude scales 
was removed which resulted in the shifts 
$\Delta X$
of the amplitudes 
$A_{00}^{(X,{\rm rev})}=A_{00}^{(X)}+\Delta X$
in Eq. (2) of \citet{barc1} where
$\Delta U=+0.020$,
$\Delta B=+0.010$,
$\Delta V=+0.049$,
$\Delta R_C=+0.029$,
$\Delta I_C=-0.017$.
The revised colours 
$U,B,R_C,I_C$ 
were interpolated to the epoch of  
$V$ 
and the colour indices were obtained by subtractions. 
The averages of the revised colour indices
are now in better agreement with those of \citet{ibvs1}. 
The observations of 
${\rm HJD}-2454200=17.3765\;
20.42\mbox{-}.47,\;44.34\mbox{-}.38,\;45.3464,45.3808$
\citep{barc1}
were omitted because of poor weather conditions. Our final list
contains  
$V,B-V,U-B,V-R_C,V-I_C$
points for 
$N=529$ 
epochs. These are available in electronic 
form{\footnote{http://www.konkoly.hu/staff/benko/pub.html}.

\begin{table}
  \caption{Reddening, metallicity, surface gravity, effective
           temperature and angular radius of the stars. The errors 
	   are
           $\Delta E(B-V)=\pm 0.01$,
           and
           $\Delta \rm{[M]}=\pm 0.05$
           for the comparison stars.}
\label{tab3}
\begin{tabular}{llllll}
\hline
ID        & $E(B-V)$  
        & [M] & $\log g_{\rm e}$ & $T_{\rm e}$ 
	& $\vartheta\times 10^{11}$  \\
	& [mag] & [dex] & [cms$^{-2}$] 
	& [K] & [rad] \\
\hline
GSC 4868 &&&& \\
$-0063$ & $0.01$  & $-0.09$  & $3.98$ &  $6550$ & $5.05$  \\
$-0779$ & $0.00$  & $+0.20$  & $4.07$ &	 $6284$ & $5.31$  \\
$-0860$ & $0.01$  & $+0.70$  & $5.20$ &  $6880$ & $4.92$  \\
$-1089$ & $0.01$  & $-0.15$  & $3.08$ &  $6148$ & $4.90$  \\
$-0831^\dag$ & $0.008$  & $-1.05$ & $3.69^\ast$ & $6902^\ast$ 
	& $10.17^\ast$ \\
	& $\pm 0.002$ & $\:\pm 0.10$  & $\:0.50$ & $\:\:262$ 
	& $\:\:\:0.32$ \\
GSC 5002 &&&&& \\
$-0506$ & $0.005$ & $-0.10$  & $4.03$ &  $6247$ & $8.82$ \\
$-0566$ & $0.01$  & $-0.05$  & $3.83$ &  $6228$ & $4.62$  \\
V372 Ser$^\ddag$ & $0.003$ & $-0.53$ & $3.24^\ast$ & $6713^\ast$ 
	& $8.13^\ast$ \\
	& $\pm 0.003$ & $\:\pm 0.05$  & $\:0.36$ & $\:\:323$ 
	& $\:0.16$ \\
\hline
\end{tabular}
{\it Notes.}
\begin{list}{}{}
\item[$^\dag$] $N^{\rm (I)}=39$. The colour curve segments
HJD-2454800=71.6077-71.6220,73.5203-73.5919,76.4228-76.5088
were used in Eqs. (\ref{3.10},\ref{3.11}).
\item[$^\ast$] Averaged values of $\log g_{\rm e},T_{\rm e},\vartheta$
from all observations. In the next row, the estimated standard errors of
$E(B-V)$,
[M] 
and the standard deviations of 
$\log g_{\rm e}$, 
$T_{\rm e}$
and
$\vartheta\times 10^{11}$
are given.
\item[$^\ddag$] $N^{\rm (I)}=249$, the colour curve segments
HJD-2454200=17.3765-17.5938, 
23.3745-23.5858, 42.3402-42.5119, 45.5006-45.6265,
48.3929-48.4986, 51.4163-51.6057
were used in Eqs. (\ref{3.10},\ref{3.11}).
\end{list}
\smallskip

\end{table}

\begin{figure*}
  \includegraphics[width=126mm]{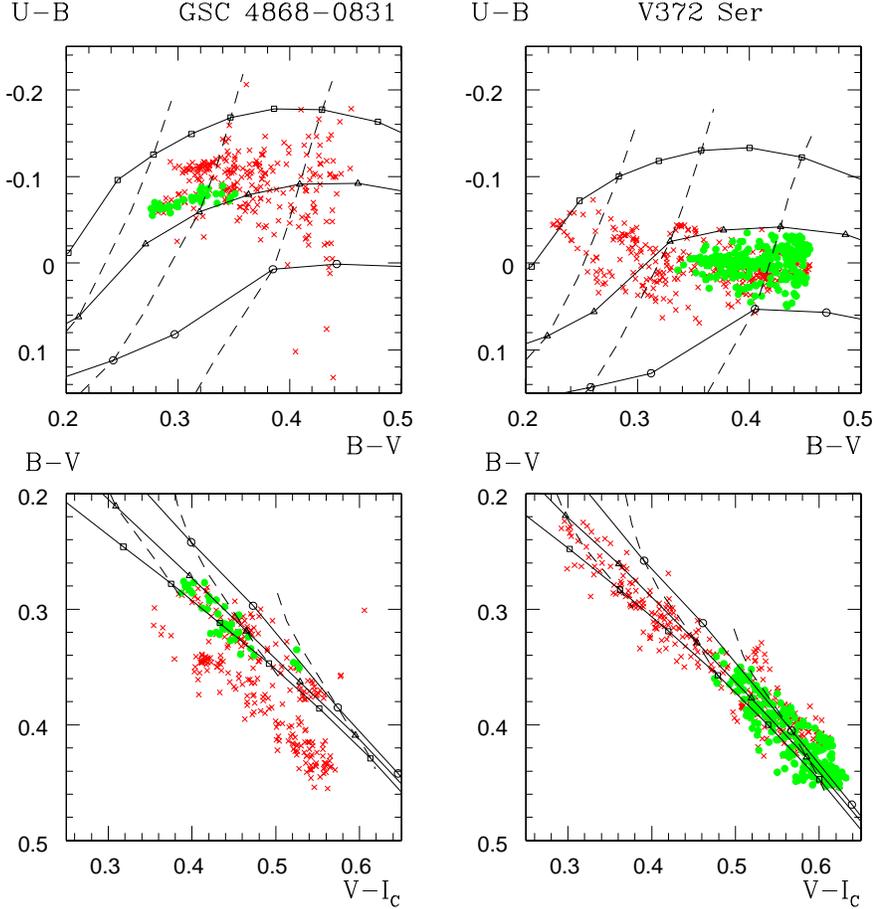}
\vspace{-0.5cm}
\caption{Colour-colour diagrams
  $(U-B)$-$(B-V)$
  and
  $(B-V)$-$(V-I_C)$.
  The panels on the 
  left and right show GSC 4868-0831 and V372 Ser, respectively.
  The solid green circles and red crosses denote that 
  ${\rm C}^{\rm (I)}$ is satisfied
  and is not satisfied, respectively. The 
  iso-$T_{\rm e}$ and
  iso-$\log g_{\rm e}$
  lines of the ATLAS models \citep{kuru1} are interpolated to
  $E(B-V)$
  and
  [M]
  as given in Table~\ref{tab3}. 
  The lines with open squares, triangles, circles denote
  $\log g_{\rm e}=4.5,3.5,2.5\;{\rm cms}^{-2}$, respectively.
  The dashed lines from left to right denote
  $T_{\rm e}=7500,7000,6500$~K.
  }
\label{fig1}
\end{figure*}

\section{Metallicity, reddening}

The physical quantities, especially
$\log g_{\rm e}$,
derived from photometry depend on 
[M]
and
$E(B-V)$.
[M] is identical with the parameter of the ATLAS models 
\citep{kuru1}. It is given on the solar scale  
(i.e. ${\rm [M]}=0$~dex for the solar composition).
Various methods to determine 
[M]
and
$E(B-V)$
are summarized in
\citet{liuj1}, none of which is applicable for our stars
except for the upper limit of
$E(B-V)$, 
which can be found from the maps of
\citet{burs1} and from the diffuse infrared 
background\footnote{{\tt http://irsa.ipac.caltech.edu/applications/DUST}} 
(DIRBE, \citealt{schl1}). The precise values are of primary importance 
because we use
$g_{\rm e}$,
$T_{\rm e}$
in the Euler equation of hydrodynamics. Therefore,
we have to use the photometric variation method described in 
\citet{barc3} and Paper I:
[M]
and
$E(B-V)$
are found from
the best fit of the observed and theoretical colour indices 
of the ATLAS models \citep{kuru1}.

When observing variable stars, we have a sum of the 
random scatter and temporal physical change in the colour indices, 
the effect of the temporal physical change manifests itself in 
increasing 
$\Delta\log g_{\rm e}$
and
$\Delta T_{\rm e}$.
To minimize this source of systematic error,
$N^{\rm (I)}$ 
phases must be selected when the photometric condition 
${\rm C}^{\rm (I)}$ 
of the QSAA is satisfied. The averages 
\begin{eqnarray}\label{3.10}
\langle\Delta\log g_{\rm e}\rangle_{N^{\rm (I)}}
	&=&{1\over{N^{\rm (I)}}}\sum_{j=1}^{N^{\rm (I)}}\Delta\log g_{{\rm e},j}, \\
\label{3.11}
\langle\Delta T_{\rm e}\rangle_{N^{\rm (I)}}
	&=&{1\over{N^{\rm (I)}}}\sum_{j=1}^{N^{\rm (I)}}\Delta T_{{\rm e},j}
\end{eqnarray}
must be minimized as a function of 
[M]
and
$E(B-V)$.

The application is straightforward for the 
comparison stars by setting  
$N^{\rm (I)}=1$
and
$\log g_{\rm e}=\log g$
because their atmosphere is static and the observed
colour indices have a random scatter only.
For orientation, 
$E(B-V)$,
[M],
$T_{\rm e}$,
$\log g$, 
$\vartheta$
of comparison stars of approximately the same
colours were determined using our method. The minimal value of 
the scatters
$\Delta\log g$,
$\Delta T_{\rm e}$
shows the effect of the random error of the colour indices. 
The averages for the different comparison stars suggest that 
the limits of the applicability of the QSAA are
$\Delta T_{\rm e} \la 5$
and
$\Delta\log g \la 0.04$
in our observational material. 
That is, 
$\mbox{C}^{\rm (I)}$
is satisfied in the phases where
$\Delta T_{\rm e}$
and
$\Delta\log g_{\rm e}$
do not exceed these limits.

The numerical results are   
$\langle \Delta\log g_{\rm e}\rangle_{39}=0.022\pm 0.001$
$\langle \Delta T_{\rm e}\rangle_{39}=8.6\pm 0.5$~K
for GSC4868-0831,
$\langle\Delta\log g_{\rm e}\rangle_{249}=0.031\pm 0.002$,
$\langle\Delta T_{\rm e}\rangle_{249}=6.2\pm 0.2$~K
for V372 Ser. The
$\Delta T_{\rm e} \la 5$~K
limit from the comparison stars is exceeded slightly. The excess 
originates from the inclusion of phases into
$N^{\rm (I)}$,
which satisfy
$\Delta\log g_{\rm e} \la 0.04$
but violate
$\Delta T_{\rm e} \la 5$~K
because of the temporal neighbourhood of an atmospheric shock.
The results are summarized in Table~\ref{tab3}. The atmospheres 
of both variables are moderately metal deficient. 

\subsection{The observed $U-B$, $B-V$, $V-I_C$ colour indices}

The observed 
$(U-B)$-$(B-V)$, 
$(B-V)$-$(V-I_C)$ 
colour-colour diagrams are plotted in 
Fig.~\ref{fig1}. The phases are plotted separately when
${\rm C}^{\rm (I)}$ is or is not satisfied (green circles and red 
crosses, respectively). The theoretical 
colour-colour relations
of the ATLAS models \citep{kuru1} belonging to some characteristic 
$\log g_{\rm e}$
and
$T_{\rm e}$ 
were interpolated for the
[M]
and
$E(B-V)$
values of the variables. These are the lines in Fig.~\ref{fig1}. Before
discussing the hydrodynamic details, we can draw some conclusions from the
data plotted in Fig.~\ref{fig1}.

\begin{itemize}
\item[$\bullet$]
The segregation of the colour indices satisfying ${\rm C}^{\rm (I)}$
is clearly seen. However, a remarkable dichotomy is obvious:
the phases of GSC 4868-0831 satisfying 
${\rm C}^{\rm (I)}$ 
are concentrated in the blue
$B-V$
domain, while those of V372 Ser populate the red
$B-V$
region.

\item[$\bullet$] Within the observational scatter, the 
$(B-V)$-$(V-I_C)$
values of V372 Ser are in the domain of
$2.5 < \log g_{\rm e} < 4.5$.
A considerable number of 
$(B-V)$-$(V-I_C)$
pairs of GSC 4868-0831, which do not satisfy
${\rm C}^{\rm (I)}$,
are significantly below this domain indicating
$\log g_{\rm e} \gg 4.5$.
A combination of
[M]
and
$E(B-V)$
could not be found that would have shifted every pair into a common
domain of
$\log g_{\rm e}, T_{\rm e}$,
[i.e. 
${\rm C}^{\rm (I)}$
is more strongly violated by the atmosphere of GSC 4868-0831].

\item[$\bullet$] As indicated by the difference in
$\overline{\log g_{\rm e}}=3.69,3.24$
(Table~\ref{tab3}),
the surface gravity of GSC 4868-0831 must be by a factor 
$\approx 3$ 
larger than that of V372 Ser, [i.e. GSC 4868-0831 is a subgiant 
rather than a giant star]. 

\end{itemize}

\section{Distance, mass and atmospheric kinematics}

The dynamical equation of the pulsation of an atmosphere with
spherical symmetry was derived in Paper I
from the Euler equation of hydrodynamics.
Its most convenient form is 
\begin{equation}\label{4.100}
{{\cal M}_{\rm a}\over{d^2}}={{\vartheta^2(t)}\over{G}}
                     \Bigl[g_{\rm e}(t) 
                     -a(r,t)-a^{\rm (dyn)}(r,t)\Bigr],  
\end{equation}
where
$a(r,t)=\partial v/\partial t+v(\partial v/\partial r)$, 
$G$
is the Newtonian gravitation constant. The dynamical correction
$a^{\rm (dyn)}(r,t)$
accounts for the difference between the accelerations of a static and
a dynamical model atmosphere at time
$t$.
The solution of the
continuity equation for mass conservation in the frames of QSAA 
resulted in a series expansion
\begin{equation}\label{4.101}
v(r,t)={\dot\vartheta}d-{1\over{h_0}}{{\partial h_0}\over{\partial t}}r
	+\cdots, \:\: r\la R.
\end{equation}
This was given in Paper I with detailed explanations for the symbols
in Eqs. (\ref{4.100}), (\ref{4.101}).
$R$
is the stellar radius [i.e. the radius of approximately zero
optical depth], 
$\vartheta=R/d$,
the dot denotes a differentiation with respect to 
$t$,
$h_0=\mu g_{\rm e}(t)/{\cal R} T(R,t)$
is the reciprocal barometric scaleheight at
$r\approx R$.

The velocity profile (\ref{4.101}) was introduced in Eq. (\ref{4.100}) 
for each epoch
$j$,
$j=1,...,N$
and 
${\cal M}_{\rm a}$
and 
$d$
were determined in four steps. In comparison with the
technique in Paper I,
the refinement of the solution of Eq. (\ref{4.100}) was rendered
possible by the large number of the observed points. 
($N=280$ and $529$
for GSC 4868-0831 and V372~Ser, respectively.)

\begin{itemize}
\item[(i)]
The photometric inverse problem [i.e. the conversion of the 
$UBV(RI)_C$
observations to 
$\vartheta,\log g_{\rm e}$
and
$h_0$]
was solved for 
$N$ 
observations, as described in Paper I, and the
$N^{\rm (I)}$
phases were found that satisfied
${\rm C}^{\rm (I)}$.  
Polynomial fits of degree 7-10 were calculated for
$\vartheta(t),\log g_{\rm e}(t), h_0(R,t)$.
One polynomial was sufficient for the whole time interval (of length
$ < 0\fd 3$)
if the atmosphere was in a shock-free state; 
$\le 3$ 
polynomials were necessary in a strongly shocked phase.

\smallskip
\item[(ii)]
Epochs of number
$n$ 
could be selected by differentiation of
$\vartheta(t)$
when
$v(r,t)/d={\dot \vartheta}+\cdots \approx \mbox{constant}$,
and the angular acceleration 
$a^{\rm (ang)}(r,t)=a(r,t)/d
		={\ddot\vartheta}
		-{\dot\vartheta}h_0^{-1}(\partial h_0/\partial t)
		+\cdots\approx 0$ 
over the whole atmosphere because the leading terms in
$v/d$
and  
$a/d$
are large in comparison to the rest. We can well assume that 
${\rm C}^{\rm (II)}$
was satisfied in these shock-free intervals
if the atmosphere was in free fall [i.e.
$a^{\rm (dyn)}(r,t)\approx 0$
holds in their vicinity].
This is the descending branch in the
light curve. Eq. (\ref{4.100}) reduces, in these epochs, to
\begin{equation}\label{4.110}
({\cal M}_{\rm a}d^{-2})_i
		=G^{-1}g_{\rm e}(t_i)\vartheta^2(t_i), 
		\: i=1,...,n,  
\end{equation} 
$n$
is given in Table~\ref{tab5}. These
$n$
epochs form a subset of the 
$N^{\rm (II)}$
epochs satisfying 
$\mbox{C}^{\rm (II)}$.

We remark that 
$a^{\rm (dyn)}(r,t)\approx 0$
might not be expected in the minimum and ascending branch of the light 
curve because there are atmospheric
layers moving in opposite direction [i.e. the atmosphere
is not in free fall]. The approximation (\ref{4.110}) is not valid 
in these phases in spite of
$a^{\rm (ang)}\approx 0$ 
for a short time when the rapidly changing 
$a^{\rm (ang)}(r,t)$
has a sign change. These shocked phases
were therefore not included in 
$i=1,2,...,n$.

The right hand side of Eq. (\ref{4.110}) consists of quantities 
derived directly from the 
photometry without differentiations of the polynomial fits. 
This can be determined with an accuracy of
$\le 0.15$
from the
$UBV(RI)_C$
photometry in each 
$t_i$. 

\smallskip
\item[(iii)]
By introducing the averaged value 
\begin{equation}\label{4.200}
{{\cal M}_{\rm a}\over{d^2}}={1\over n}\sum_{i=1}^n
		\Bigl({{\cal M}_{\rm a}\over{d^2}}\Bigr)_i
\end{equation}
we reduce Eq. (\ref{4.100}) to an algebraic equation for unknown
$d_j$
at any
$t_j, \: j=1,...,N$
if
$a^{\rm (dyn)}=0$;
the largest term is linear in
$d_j$.
These 
$N$
equations can be solved by elementary operations for all
$j$.
The distance was obtained from 
\begin{equation}\label{4.300}
d(N)={1\over N}\sum_{j=1}^Nd_j.
\end{equation}
At this step, an averaged dynamical correction
\begin{eqnarray}
\overline{a^{\rm (dyn)}}(R,N)
		&=&{1\over N}\sum_{j=1}^Na^{\rm (dyn)}(R,t_j) \nonumber \\
\label{4.310} 	&=&{1\over N}\sum_{j=1}^N\Bigl[g_{\rm e}(t_j)
		-{{G{\cal M}_{\rm a}}\over{\vartheta^2(t_j)d^2}}
		-a(R,t_j)
	\Bigr]
\end{eqnarray}
was defined with  
${\cal M}_{\rm a}d^{-2}$
and 
$d(N)$ 
from Eqs. (\ref{4.200}) and (\ref{4.300}) facilitating the reduction
of the number of the equations (\ref{4.100}) to be solved.  

\smallskip
\item[(iv)]
To obtain the final
$d(N^{\rm (II)})$
the step (iii) was repeated, reducing
$N$.      
One or two iterative steps were necessary to obtain 
$\overline{a^{\rm (dyn)}}(R,N^{\rm (II)})
		\ll \overline{g_{\rm s}[R(t)]},$
and 
$\vert a^{\rm (dyn)}(R,t_j)\vert_{j=1,...} 
	\la 0.3\overline{g_{\rm s}[R(t)]}$
where
$g_{\rm s}[R(t)]=G{\cal M}_{\rm a}R^{-2}(t)$.
That is, the roots
$d_j$
of Eq. (\ref{4.100}), originating from the 
$t_j$
intervals, were removed from Eq. (\ref{4.300}) which produced 
outlier values with
$a^{\rm (dyn)}(R,t_j)\ga 0.3g_{\rm s}[R(t_j)]$. 
The results of the iterative steps are given in  Table~\ref{tab5}.
\end{itemize}

\begin{table*}
\begin{minipage}{126mm}
  \caption{The results from steps (i)-(iv).}
\label{tab5}
\begin{tabular}{lllllll}
\hline
$\{E(B-V),[M]\}$ & $n$ & ${\cal M}_{\rm a}d^{-2}\times 10^7$ & 
	$d$ & ${\cal M}_{\rm a}$ 
	& $\overline{a^{\rm (dyn)}}(R,t)$ & $N$ \\
	${\rm [mag], [dex]}$ & 
	& [${\cal M}_{\odot}\mbox{pc}^{-2}$] & [pc]  & 
	[${\cal M}_{\odot}$] & [$\mbox{ms}^{-2}$] & \\
\hline
\multicolumn{7}{c}{GSC 4868-0831} \\
$\{0.008,-1.05\}$ & 7 & $40.3\pm 6.7$ 
	& $451\pm 99$ & $.82\pm .36$ & $-58.4$ & 280 \\ 
&& 
	& $467\pm 16$ & $.88\pm .06$ & $0.10$ & $21^\ast$  \\
\hline
\multicolumn{7}{c}{V372 Ser} \\
$\{0.003,-0.53\}$ & 20 & $6.12\pm .31$ 
	& $1183\pm 74$ & $0.85\pm .13$ & $13.90$ & 529  \\ 
&& 
	& $1020\pm 87$ & $0.63\pm .12$ & $1.64$ & 249  \\
&& 
	& $964\pm 81$ & $0.57\pm .10$ & $0.41$ & $135^\ast$  \\
\hline
\end{tabular}
\end{minipage}

{\it Note.} 
The $^\ast$ indicates
$N^{\rm (II)}$;
$\vert a^{\rm (dyn)} \vert < 3\mbox{ms}^{-2}$ 
is satisfied for all
$t_j, j=1,...,N^{\rm (II)}$.
\smallskip

\end{table*}

The interval of the search for roots of Eq. (\ref{4.100}) was limited to
\begin{equation}\label{4.108}
d_j \le d_{\rm max}=g_{\rm e}{\ddot\vartheta}^{-1}_{\rm max}.
\end{equation}
To obtain this upper limit,
we took into account only the linear term of the acceleration  
when the atmosphere is hit by the strongest shock wave; this is the 
state of maximal compression. At this epoch the atmosphere 
is just reversed from inward to outward motion. That is,
$v(\partial v/\partial r)\approx 0$,
$\partial v/\partial t={\ddot\vartheta}d > 0$,
$R$
is minimal, the deceleration by the static gravity is negligible
($g_{\rm s} \ll \partial v/\partial t$)
and, furthermore, 
$a^{\rm (dyn)} \approx 0$
was supposed. However, the upper limit
$d_{\rm max}$
was taken into account only if the satisfaction of these conditions plus,
${\rm C}^{\rm (I)}$ and 
${\rm C}^{\rm (II)}$,
was verified afterwards. 

\subsection{Numerical results}

Our observations provided
10 light curve segments
of sufficient length 
and quality to calculate the polynomial fits of
$\log g_{\rm e}(t), T_{\rm e}(t)$
and
$\vartheta(t)$
for both stars.
The order of magnitude limits were found to be
$-10^{-20} < {\rm O}[a^{\rm (ang)}] < 10^{-18}\mbox{rad}\cdot \mbox{s}^{-2}$
for both stars. The acceleration-free intervals can be found for
Eq. (\ref{4.200}) from quantities that are of the form
$f(\vartheta)+{\rm O}(d^{-1})$. 

The results from step (i),
[M]
and
$E(B-V)$
are given in Table~\ref{tab3}.
The results from steps (ii)-(iv) and the final
$d$
and
${\cal M}_{\rm a}$
are summarized in Table~\ref{tab5}. 
The asterisk in the last column denotes the value of
$N^{\rm (II)}$.

The position of the stars in a theoretical HRD
was calculated by 
\begin{eqnarray}\label{4.901}
T_{\rm eq}&=&\langle\vartheta^2T_{\rm e}^4\rangle^{1/4}  
	\langle\vartheta\rangle^{-1/2},  		\\
\label{4.900}
L_{\rm eq}&=&4\pi\sigma d^2
	\langle\vartheta^2(\varphi) T_{\rm e}^4(\varphi)\rangle  
\end{eqnarray}
\citep{carn1}.
The numerical values are given in Table~\ref{tab7}. 
The remarkably small error of
$T_{\rm eq}$
originates from the sum of the scatter of
$T_{\rm e}$,
determined from the different colour-colour combinations and the
scatter of
$\vartheta$
when it is determined from the brightness in
$V,R_C$
and
$V$ plus bolometric correction and the Stefan-Boltzmann law.
The components of the error of
$L_{\rm eq}$
are composed of errors from
$\langle\vartheta^2T_{\rm e}^4\rangle$,
${\cal M}_{\rm a}d^{-2}$
and
$d$,
respectively. The numerical values are
$\pm 0.18,\pm 0.35,\pm 0.70$
for GSC 4868-0831
$\pm 0.2,\pm 1.3,\pm 3.7$
for V372~Ser. A summary of the parameters is given in Table~\ref{tab7}. 

The error of the magnitude averaged
$\langle M_V \rangle$
was estimated from sum of the errors of
${\cal M}_{\rm a}d^{-2}$
and
$d$ 
in Table~\ref{tab5}.

Fig.~\ref{fig3} presents an insight into the variable quantities of
the atmosphere at a shock-free and a shocked phase for both stars.
The segment of
$V(t)$
is plotted in the uppermost row for orientation. The scatter of
$T_{\rm e}(t)$ 
(row 2) reflects the scatter in our photometry.
$R(t)$
and the velocities and accelerations were computed from the smoothed
$\vartheta(t)$
and
$\log g_{\rm e}(t)$
using
$d$
and
${\cal M}_{\rm a}$
from Table~\ref{tab5}.

We see a considerable gradient of 
$v(r)$ 
for GSC 4868-0831 at HJD$\ga 2454829.6$,
just when a descending branch of the light curve ended. The deeper
layers were moving outwards much faster 
because the atmosphere was hit by a shock wave, and 
the velocity excess reached
$\approx 25$~kms~$^{-1}$. 
Otherwise
$\vert v(R)-v(R-h_0^{-1})\vert\approx$ a few km s~$^{-1}$
was found for both stars.

Row 5 is a plot of
$g_{\rm s}(R,t)$.
The different types of GSC 4868-0831 and V372 Ser are obvious 
if we compare 
$R(t)$
and
$g_{\rm s}(R,t)$;
GSC 4868-0831 is a subgiant star while V372 Ser has a true
RRd character.
The components of the acceleration are plotted in row 6 in the relative
units
$q_d=a^{(\rm dyn)}(R,t)g^{-1}_s(R,t)$,
$q_r={\ddot R}(R,t)g^{-1}_s(R,t)$,
$q_a=a(R,t)g^{-1}_s(R,t)$.
Here, 
$\vert q_d \vert \la 0.3$
is the domain when
${\rm C}^{\rm (II)}$
is satisfied if the light curve is not in ascending branch. The rapid
changes of the components were not smoothed out, especially those of
$q_d$
of V372 Ser at
HJD$> 2454250.57$. These originate partly from violating
${\rm C}^{\rm (I)}$,
and partly from the stronger effect of the thermic shock on the 
more dilute atmosphere of V372 Ser. 
$q_d \ga 10$
means that 
$g_{\rm e}$
derived from the photometry has a very loose connection with the 
atmospheric kinematics. This must be corrected by a term
$\gg g_{\rm s}$
to obtain the actual acceleration in the atmosphere.

\begin{figure*}
  \includegraphics[width=170mm]{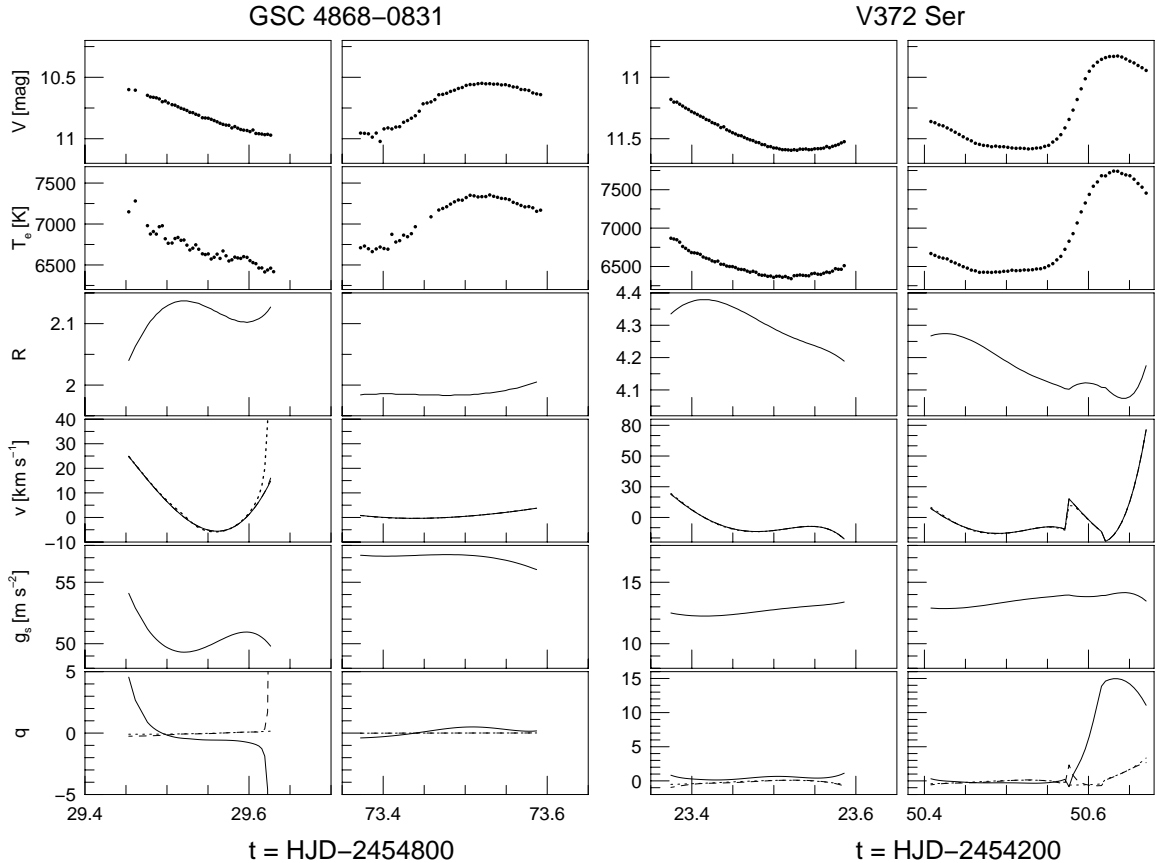}
\vspace{-0.8cm}
\caption{The variable physical parameters of GSC 4868-0831 and V372 Ser
in descending and ascending branches of the 
$V$
light curve. 
Row 1: light curves
	$V(t)$.
Row 2: 
	$T_{\rm e}(t)$.
Row 3:  $R(t)$.
Row 4: solid lines:
	$v(R,t)$,
	dotted lines:
	$v(R-h_0^{-1},t)$.
Row 5: $g_{\rm s}(R,t)$.
Row 6: relative accelerations
	(solid lines:
	$q_d=a^{(\rm dyn)}(R,t)g^{-1}_s(R,t)$,
	dotted lines:
	$q_r={\ddot R}(R,t)g^{-1}_s(R,t)$,
	dashed lines:
	$q_a=a(R,t)g^{-1}_s(R,t)$).
}
\label{fig3}
\end{figure*}

\section{Discussion}

\subsection{Remarks on the photometric inverse problem}

Because of the lack of high-dispersion spectra the
relations
\begin{equation}\label{5.100}
{\cal P}=f({\rm CI}_1,{\rm CI_2}...)
\end{equation}
play a crucial role. Here,
${\cal P}$
and 
$\mbox{CI}_i$
denote an astrophysical quantity and colour index, respectively.
For example, a particular form of
$f$
belongs to 
${\cal P}=T_{\rm e}$,
${\cal P}=\log g$,
etc.  These are given in tabular form for the ATLAS models 
\citep{kuru1}, and their parameters are 
$E(B-V)$,
[M], etc.
(As emphasized in Sec. 3, 
[M]
is on the solar scale and
$f$
would have another form for a peculiar atmospheric chemical 
composition.)

The derived fundamental parameters depend on the reddening and 
metallicity. A larger
$E(B-V)$
leads to larger
$\log g$
and 
$T_{\rm e}$,
the dependence on
[M]
is similar, but to a lesser extent.
 The essence of our method 
is that we search for the minimal standard errors
$\Delta\log g_{\rm e},\Delta T_{\rm e}$
of
$\log g_{\rm e},T_{\rm e}$
from some 30 colour index pairs as a function of 
[M] 
and
$E(B-V)$. This method is self-consistent
to the highest degree from an astrophysical point of view. We use 
all colour information of the atmospheric models to determine 
the relevant parameters and quantities. 
In doing so, we can avoid the systematic errors
inherent to an arbitrary choice of relations such as, for example,
$T_{\rm e}$-$(V-I_C)$,
${\rm [M/H]}$-$(B-V)$,
(e.g. \citealt{deka1}).
In addition, we do not use semi-empirical calibrations
at all (e.g. the $T_{\rm e}$-colour index relations derived
for non-variable main sequence stars of normal chemical composition 
and, as next step, generalized to variable giant stars with large
metal deficiency, \citealt{clem1}). 

Of course, in comparison with a line by line analysis of
high-dispersion spectroscopy, a multicolour photometry
is sensitive only to the general shape of the optical flux of the star.
However, it is possible to find the model atmosphere by reproducing
the continuum flux in the visible wavelength interval. Our method
can provide a global parameter of the chemical composition in the
atmosphere summarized as metallicity
[M].
This [M] is the most suitable for our purposes, because it accounts 
for the effect of all elements on the optical continuum, including those
that do not have lines. Of course, it 
might differ from the overall metallicity of the star (used in the
theory of stellar structure, as the 
pulsation does not stir up the deepest layers where the nuclear
reactions take place) or from the averaged metallicity derived from
high dispersion spectroscopy.

The fields are at high Galactic latitude: 
GSC 4868-0831, $b= +24\degr$; V372 Ser, $b=+45\degr$.
Their reddening is very small, and therefore the parameters
($\log g_{\rm e}$, 
$T_{\rm e}$, 
$d$, 
${\cal M}_{\rm a}$)
reported in Section 4 are lower limits at the same time if
[M] 
is fixed. The upper limits are
$E(B-V) < 0.023, 0.085$
\citep{schl1},
or
$E(B-V) < 0.03, 0.045$
\citealt{burs1}, respectively. We attribute our
smaller 
$E(B-V)$
to two factors: 
our method measures the reddening of a point source and
the excess reddening must originate from a region beyond 
GSC 4868-0831, V372 Ser, and the comparison stars.

The reddening
$E(B-V)=0.085$
of V372 Ser derived from the DIRBE differs from our value above the
$5\sigma$
level. Taking this large 
$E(B-V)$ 
would result in an unacceptable increment from
$\langle \Delta T_{\rm e}\rangle_{249}=6.2\pm 0.2$ 
to
$8.2\pm 0.4$.
The parameters would increase to 
$\overline{\log g_{\rm e}}=4.11$,
$\overline{T_{\rm e}}=7092$
leading to a mass above
$5{\cal M}_\odot$
which cannot be reconciled with any actual theoretical knowledge about
pulsating stars. Therefore, 
$E(B-V) > 0.006$
can be ruled out.

\subsection{The use of $U$ observations}

Concerning GSC4868-0831, 
an interesting result can be seen from Fig.~\ref{fig1}: the phases 
satisfying and not satisfying
${\rm C}^{\rm (I)}$
segregate clearly in the colour-colour diagrams and completely different
$T_{\rm e}(t)$
and 
$\log g_{\rm e}(t)$
are obtained for these phases if a
$BVI_C$
photometry only is used as input.  
Fig.~\ref{fig1} demonstrates that reliable 
$T_{\rm e}(t)$ 
and 
$g_{\rm e}(t)$
can be determined only if colour indices containing 
$U$
are used in the
$T_{\rm e}$,
$\log g_{\rm e}$
domain of RR stars.
The use of one colour index or 
$BVI_C$
photometry only is not sufficient and can be misleading,
even if it is limited to determining
$T_{\rm e}$
only. 

Furthermore, our
$U$
observations might reveal the subgiant character of GSC 4868-0831
which was not suspected previously (\citealt{wils3},
\citealt{wils2}, \citealt{szcz1}).
An important conclusion has emerged that it is not possible
to determine the luminosity class of a pulsating star if only periods
or period ratio are available. Multicolour observations, covering the
ultraviolet, are needed to classify DM pulsators properly. 

\subsection{Remarks on the QSAA}

We have to emphasize that 
$L_{\rm eq}$
and
$T_{\rm eq}$
are first approximations from Eqs. 
(\ref{4.901}) and (\ref{4.900}) because some elements of the averaging
were obtained assuming QSAA in phases when 
${\rm C}^{\rm (I)}$ 
was violated. Qualitative considerations suggest a positive 
correction to
$T_{\rm e}(t)$
of QSAA in the shocked phases when excess
radiation and dissipation exist from shock waves. 
Corrections emerging from a dynamical model 
atmosphere would not modify the main fundamental parameters
${\cal M}_{\rm a}$
and
$d$
because they were determined from phases when
both quantitative conditions of the validity of QSAA were satisfied.
${\cal M}_{\rm a}$
and
$d$
can be considered as well substantiated empirical data from the 
ATLAS static model atmospheres plus some
basic hydrodynamics. Dynamical model atmospheres are beyond the
scope of this series of papers. 

The sampling of the quasi-repetitive curves introduced negligible
error, which can be estimated by comparing 
$T_{\rm e}$
and
$\vartheta$
from the fitted colour curves \citep{barc1} with those from the
$N=529$
observations of V372 Ser.

We remark that 
$d=(1145\pm 73)$~pc,
${\cal M}_{\rm a}=(0.83\pm .17){\cal M}_\odot$
are the results for
$\{E(B-V)=0.003,[M]=-0.53\}$
if the UAA is applied, that is, if
$\partial v/\partial r=0$
is assumed in Eq. (\ref{4.100}) (and, as a consequence, 
$\vartheta=R/d$
and
$\partial v/\partial t={\ddot\vartheta}d$).
This distance and the change of
${\cal M}_{\rm a}d^{-2}=6.12\pm .31 \rightarrow 7.64\pm .82$
gives
$L_{\rm eq}^{(\rm UAA)}=42.8L_\odot$
which put V372 Ser in a position just at the lower limit of stable
DM pulsation \citep{szab1}.  
Of course, the physical input of the UAA is much less than that of our
extended hydrodynamic treatment represented by Eqs. (\ref{4.100}), 
(\ref{4.101}). In spite of the better agreement,
the data from UAA must not be accepted because the UAA is a rigid and less
realistic approximation in comparison with a compressible model atmosphere. 

\begin{figure}
  \includegraphics[width=84mm]{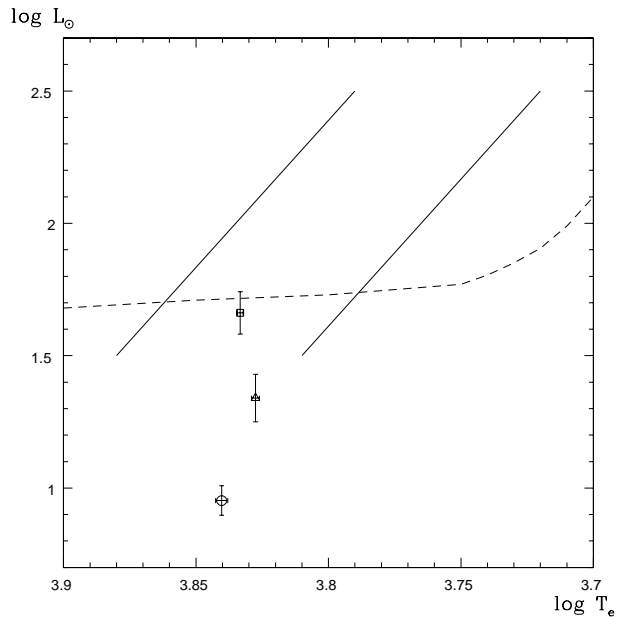}
\vspace{-0.8cm}
\caption{The positions of the stars in a theoretical HRD.
The square denotes SU Dra,
(with $[L_{\rm eq},T_{\rm eq}]$
were taken from Paper I), 
The triangle and circle denote V372 Ser and GSC 4868-0831, respectively.
The solid lines from left to right denote the blue and red edges 
of the instability strip.
The dashed line denotes the zero age horizontal branch of M3 
\citep{agui1}. 
  }
\label{fig4}
\end{figure}

\subsection{Classification of the stars}

The positions of our stars and SU Dra in a theoretical HRD, 
(i.e. 
$[L_{\rm eq},T_{\rm eq}]$)
are plotted in Fig.~\ref{fig4}. For orientation, the instability 
strip and the zero age horizontal branch of the
metal deficient 
([M]$\approx -1.6$)
globular cluster M3 (NGC 5272) are shown \citep{agui1}. We emphasize 
that 
$[L_{\rm eq},T_{\rm eq}]$
correspond to
$L$
and
$T_{\rm e}$
of non-variable stars, and they
can be directly compared with
those from the theoretical studies on stellar structure, pulsation
and evolution. The only source of error is the difference of
$T_{\rm e}(t)$
and
$\vartheta(t)$
in Eqs. (\ref{4.901}, \ref{4.900})
from QSAA and dynamical model atmospheres, respectively.
An error has not been propagated into the 
position of the star by semi-empirical relations like
$\overline{T_{\rm e}}$-$\overline{(B-V)}$
etc. 

The period ratios 
$P_1/P_0$
are
$=0.7443$
and
$0.7450$
for V372 Ser and GSC 4868-0831, respectively. These
are in the canonical range of RRd stars, as 
${\cal M}_{\rm a}$
of V372 Ser is also.
The equilibrium effective temperatures of both stars are in the 
$5630 < T_{\rm eq} < 7080$~K 
interval where DM pulsation can be stable. 
The results of our analysis confirm that V372 Ser is a true 
RRd star. However, it is subluminous by a factor of
$\approx 2\mbox{-}3$
in comparison with the pulsation models of RRd stars \citep{szab1}.

${\cal M}_{\rm a}$ of GSC 4868-0831 exceeds somewhat the 
canonical mass of RRd stars; its subluminosity is even larger,
$\approx 4\mbox{-}5$
and its radius is less than half that of an RR star. The periods
exclude its identification as DM SX Phe star.
To the best of our knowledge, GSC 4868-0831 is the first known
subgiant star pulsating in two modes of large amplitude. It is 
slightly 
($\approx 1$~mag)
above the zero age main sequence of normal chemical composition,
and it is below the extension of the instability strip of M3.

\section{Conclusions}

We have reported 302
$UBV(RI)_C$
observations of GSC 4868-0831. To the best of our knowledge, this is
the second DM pulsating star with well documented 
photometric behaviour in the optical and near ultraviolet bands.  

We have determined, for the first time, fundamental parameters of
DM pulsators using only the theory of stellar atmospheres. 
Our method is purely photometric; the eventual uncertainty of 
observing radial
velocities and their conversion into the reference frame in the
centre of the star can be avoided using our method. Our method
has made extensive use the ATLAS model atmospheres and some
basic hydrodynamics. Our astrophysical input is completely
different from the theory of stellar pulsation and evolution.
Therefore, our method can be used as a check or challenge for 
these more involved theories.

We have summarized our results in Table~\ref{tab7}. 
We have found the surprising result that GSC 4868-0831 is not an RRd
star, but a subgiant star pulsating in two modes. 
We have found a significant 
subluminosity of the RRd star V372 Ser in comparison with the luminosity 
from present day theory of stellar pulsation.
These objects must be considered as a challenge to extend the search for
stable DM pulsation.
Qualitative considerations suggest that dynamical model atmospheres
would have higher 
$T_{\rm eq}$
and
$L_{\rm eq}$;
both corrections would shift the points in Fig.~\ref{fig4} upwards
and to the left with respect to their position from QSAA. 
However, we must not forget that
our results from static ATLAS model atmospheres have yielded 
the ratio
${\cal M}_{\rm a}d^{-2}$
and
$d$
from phases when a dynamical model atmosphere is not necessary (i.e. 
QSAA is a reliable approximation of the pulsating atmosphere). 
We are faced with the dilemma that there is either  
subluminosity with acceptable mass or acceptable luminosity 
with too large mass if we are to reconcile these parameters from the
present-day theory of stellar pulsation.

\begin{table}
	\caption{Summary of the parameters.}
\label{tab7}
\begin{tabular}{lll}
\hline
	& GSC 4868-0831 & V372 Ser \\
\hline
$P_0$  & $0\fd 5649^\ddag$ & $0\fd 4712289^\dag$ \\ 
$A_0$  & $0.07271^\ddag$ mag & $0.1534^\dag$ mag \\
$P_1$  & $0\fd 42085^\ddag$ & $0\fd 3507310^\dag$ \\ 
$A_1$  & $0.15815^\ddag$ mag & $0.2059^\dag$ mag \\
$E(B-V)$ & $0.008\pm .002$ mag & $0.003\pm .003$ mag \\
${\rm [M]}$ & $-1.05\pm .10$ dex & $-0.53\pm .05$ dex\\
$d$ & $467\pm 16$ pc & $964\pm 81$ pc \\
$\langle M_V \rangle$	& $+2.40\pm .24$ mag & $+1.58\pm .22$ mag \\
${\cal M}_{\rm a}$ & $0.88\pm .06 {\cal M}_\odot$ & $0.57\pm .10{\cal M}_\odot$ \\
$R_{\rm min}$	& $1.97R_\odot$	& $4.07R_\odot$	\\
$R_{\rm max}$	& $2.05R_\odot$	& $4.40R_\odot$	\\
$T_{\rm eq}$	& $6924\pm 35\mbox{K}$ & $6722\pm 20\mbox{K}$	\\
$L_{\rm eq}$	& $8.97\pm 1.23L_\odot$	
		& $21.9\pm 5.2L_\odot$	\\
\hline

\end{tabular}

Notes. 

$^\dag$: From \citet{barc1}.

$^\ddag$: determined from the $V$ observations.

\smallskip
\end{table}

\section*{Acknowledgements}

We are grateful for the travel support by the Hungarian 
Astronomical Foundation and for the hospitality at the Teide
Observatory, IAC. We have used the SIMBAD data of CDS.
We are grateful to R. Szab\'o for reading the text and improving the
English. We thank the referee, J. Nemec, for valuable
remarks and constructive suggestions.

\appendix
\section{Description of the program package BBK}

The package BBK is designed to extract the fundamental
parameters of RR stars from high-quality
$UBV(RI)_C$
observations. A brief description and some technical details
are given here.

It is important to have the colours and colour
indices as close as possible to the international colour system 
which was applied by \citet{kuru1} in order
to convert the physical fluxes of the ATLAS models to the stellar 
magnitude and colour system. An error (or errors) in the colour
indices can lead to false results. To obtain reliable results,  
some 
$N \ga 300$ 
five colour observations are needed. These must be distributed 
uniformly over a representative light curve because the physical 
quantities 
$\vartheta$,
$h_0$
must be differentiated. 

Proper use of the program package requires basic knowledge of
absolute stellar photometry, the theory of stellar atmospheres and
hydrodynamics. As \citet{kuru1} said: 'Neither the programs nor data 
are black boxes.
You should not be using them if you do not have some understanding
of the physics and of the programming in the source code.' 
This warning is appropriate for BBK.  

BBK consists of FORTRAN programs, z-shell scripts, input files,
data files, and a manual \citep{barc7}.  
UNIX or LINUX environment, installation of a
FORTRAN compiler, z-shell, graphical packages (e.g. SUPERMONGO, 
or GNUPLOT) are necessary. Five steps are involved, a
detailed description of which can be found in the manual. 
Inspection, evaluation, plots of the partial results 
are necessary before continuing to the next step. 

\vspace{3pt}

\noindent
{\sl Step (I)}
The atmospheric metallicity and the 
interstellar reddening of the star are determined from selected phases 
when the atmosphere is free of shocks (i.e. from colour indices
in the descending branch).
\vspace{3pt}

\noindent
{\sl Step (II)} 
The conversions
\begin{equation}\label{A1.100}
f({\rm CI}_1,{\rm CI_2}...) \rightarrow {\cal P}
\end{equation}
are executed to select the static ATLAS models with the best fit
to the observed and theoretical colour indices to obtain
$\log g_{\rm e}$
and
$T_{\rm e}$
as a function of phase. 
The variation of
$\vartheta$
is determined by comparing the
physical fluxes of the star with those of the selected theoretical
models. The physical fluxes 
$V$,
$R_C$,
$V+$~bolometric correction are used, those of the star are
calculated from the absolute calibration of Vega \citep{tug1}.)  
\vspace{3pt}

\noindent
{\sl Step(III)} 
Polynomial fits are calculated to obtain 
$\log g_{\rm e}$,
$T_{\rm e}$,
$h_0$,
angular velocity and acceleration 
${\dot\vartheta},{\ddot\vartheta}$
as a function of phase.
The upper limit
$d_j \le d_{\rm max}=g_{\rm e}{\ddot\vartheta}^{-1}_{\rm max}$
must be fixed by inspecting
${\ddot\vartheta}$
as a function of phase.
\vspace{3pt}

\noindent
{\sl Step(IV)} 
The fits are introduced in the Euler equation of hydrodynamics
to determine the angular acceleration and to
find the acceleration-free phases of the atmosphere.
The transient acceleration-free
intervals in the ascending branch (i.e. in the shocked phases)
must be manually deleted from the averaging to find
${\cal M}_{\rm a}d^{-2}$.
\vspace{3pt}

\noindent
{\sl Step(V)} 
$\vartheta$,
${\dot \vartheta}$,
${\ddot\vartheta}$,
$h_0$,
$\partial h_0/\partial t$,
$g_{\rm e}$
and
${\cal M}_{\rm a}d^{-2}$
are introduced in Eqs. (\ref{4.100},\ref{4.101}) and Eq. (\ref{4.100})
is solved for
$d$
with the whole (or partial) set containing 
$N$
(or
$<N$)
phase points. The upper limit 
$a^{\rm (dyn, \; upper\; limit)}$
must be specified to exclude the phase points from the 
final solution for
$d$
[i.e. from
$N^{\rm (II)}$]
to achieve 
$\vert a^{\rm (dyn)}\vert < a^{(\rm dyn,\; upper\; limit)}$
in all phase points giving
$d$.
One or two runs might be necessary with decreasing
$N$.

\end{document}